\documentclass[prl,reprint,aps,tightenlines,preprintnumbers,amsmath,amssymb,
showpacs,superscriptaddress]{revtex4-1}
\usepackage{amsmath,amssymb,graphicx,natbib}
\usepackage{graphicx}
\usepackage{dcolumn}
\usepackage{bm}
\usepackage[hyperindex,breaklinks]{hyperref}
\usepackage{amsmath}
\usepackage{array}

\def\be{\begin{equation}}
\def\ee{\end{equation}}
\def\ba{\begin{eqnarray}}
\def\ea{\end{eqnarray}}
\def\ge{\mathrel{\raise.3ex\hbox{$>$\kern-.75em\lower1ex\hbox{$\sim$}}}}
\def\la{\mathrel{\raise.3ex\hbox{$<$\kern-.75em\lower1ex\hbox{$\sim$}}}}

\def\simgt{\mathrel{\raise.3ex\hbox{$>$\kern-.75em\lower1ex\hbox{$\sim$}}}}
\def\simlt{\mathrel{\raise.3ex\hbox{$<$\kern-.75em\lower1ex\hbox{$\sim$}}}}

\newcommand{\fr}[2]{\frac{#1}{#2}}

\newcommand{\nc}{\newcommand}

\nc{\gone}{\bar g_{\pi NN}^{(1)}}
\nc{\gzero}{\bar g_{\pi NN}^{(0)}}
\nc{\al}{\alpha}
\nc{\ga}{\gamma}
\nc{\de}{\delta}
\nc{\ep}{\epsilon}
\nc{\ze}{\zeta}
\nc{\et}{\eta}

\nc{\Th}{\Theta}
\nc{\ka}{\kappa}
\nc{\rh}{\rho}
\nc{\si}{\sigma}
\nc{\ta}{\tau}
\nc{\up}{\upsilon}
\nc{\ph}{\phi}
\nc{\ch}{\chi}
\nc{\ps}{\psi}
\nc{\om}{\omega}
\nc{\Ga}{\Gamma}
\nc{\De}{\Delta}
\nc{\La}{\Lambda}
\nc{\Si}{\Sigma}
\nc{\Up}{\Upsilon}
\nc{\Ph}{\Phi}
\nc{\Ps}{\Psi}
\nc{\Om}{\Omega}
\nc{\ptl}{\partial}
\nc{\del}{\nabla}
\nc{\ov}{\overline}
\nc{\newcaption}[1]{\centerline{\parbox{15cm}{\caption{#1}}}}

\newcommand{\mav}{\ensuremath{\overline{m}}}
\newcommand{\hp}{\ensuremath{\mathrm{H}^+}}
\newcommand{\lip}{\ensuremath{\mathrm{Li}^+}}
\newcommand{\hydr}{\ensuremath{\mathrm{H}}}

\newcommand{\hef}{\ensuremath{{}^4\mathrm{He}}}
\newcommand{\het}{\ensuremath{{}^3\mathrm{He}}}

\newcommand{\lisv}{\ensuremath{{}^7\mathrm{Li}}}

\newcommand{\ben}{\ensuremath{{}^9\mathrm{Be}}}

\def\beq{\begin{equation}}
\def\eeq{\end{equation}}
\def\bmat{\begin{displaymath}}
\def\emat{\end{displaymath}}
\def\bear{\begin{eqnarray}}
\def\eear{\end{eqnarray}}
\def\bery{\begin{array}}
\def\ery{\end{array}}
\def\bit{\begin{itemize}}
\def\eit{\end{itemize}}
\def\ben{\begin{enumerate}}
\def\een{\end{enumerate}}
\def\btab{\begin{tabular}}
\def\etab{\end{tabular}}
\def\btbl{\begin{table}}
\def\etbl{\end{table}}
\def\bfig{\begin{figure}[htb]}
\def\efig{\end{figure}}
\def\bpic{\begin{picture}}
\def\epic{\end{picture}}

\def\ga{\mathrel{\raise.3ex\hbox{$>$\kern-.75em\lower1ex\hbox{$\sim$}}}}
\def\la{\mathrel{\raise.3ex\hbox{$<$\kern-.75em\lower1ex\hbox{$\sim$}}}}
\def\gappeq{\mathrel{\rlap {\raise.5ex\hbox{$>$}}
{\lower.5ex\hbox{$\sim$}}}}
\def\lappeq{\mathrel{\rlap{\raise.5ex\hbox{$<$}}
{\lower.5ex\hbox{$\sim$}}}}

\def\gyr{{\rm \, G\kern-0.125em yr}}
\def\mev{{\rm \, Me\kern-0.125em V}}
\def\gev{{\rm \, Ge\kern-0.125em V}}
\def\tev{{\rm \, Te\kern-0.125em V}}

\begin{document}

\title{Lithium Diffusion in the Post-Recombination Universe and Spatial Variation of [Li/H]   }

\author{Maxim Pospelov}
\affiliation{Perimeter Institute for Theoretical Physics, Waterloo, ON N2L 2Y5, 
Canada}
\affiliation{Department of Physics and Astronomy, University of Victoria, 
Victoria, BC V8P 5C2, Canada}

\author{Niayesh Afshordi}
\affiliation{Department of Physics and Astronomy,
University of Waterloo, Waterloo, ON, N2L 3G1, Canada}
\affiliation{Perimeter Institute for Theoretical Physics, Waterloo, ON N2L 2Y5, 
Canada}

\begin{abstract}

The observed amount of lithium for low metallicity population II stars (known as the Spite plateau) is a factor of $\sim 3-5$ 
lower than the predictions of the standard cosmology. Since the observations are limited to the local Universe (halo stars, globular clusters and satellites of the Milky Way) it is possible that certain physical processes may have led to the spatial 
separation of lithium and local reduction of [Li/H]. We study the question of lithium diffusion after the 
cosmological recombination in sub-Jeans dark matter  haloes, taking into account that more than 95\% of lithium remains in the singly-ionized state at all times. Large scattering cross sections on the rest of the ionized gas leads to strong coupling of lithium to protons and its initial direction of diffusion coincides with that of H$^+$. In the rest frame of the neutral gas this leads to the diffusion of H$^+$ and Li$^+$ out of overdensities with the trend of reducing [Li/H] in the minima of gravitational wells relative to the primordial value. 
We quantify this process and argue  that, with certain qualifications, it may have played a significant role in creating local lithium deficiency 
within the primordial dark matter haloes, comparable to those observed along the Spite plateau.

\end{abstract}

\maketitle

\newpage
The primordial abundances of light elements, $^4$He, D, and $^7$Li, 
offer unique window into the very early Universe at redshifts of $z\sim 10^9$. 
In recent years, this probe has been sharpened: the only free parameter that enters the standard 
big bang nucleosynthesis (SBBN) calculations -- baryon-to-photon ratio $\eta_b$ -- 
has been measured to great accuracy via the CMB experiments. The prediction for the primordial fraction of $^7$Li is 
$[{\rm ^7Li}/{\rm H}]_{\rm SBBN} = (5.07^{+0.71}_{-0.62})\times 10^{-10}$ (see {\em e.g.} \cite{Keith}),
which is a factor of $3-5$ smaller than the Spite plateau value of \lisv, 
$(1.23^{+0.34}_{-0.16})\times10^{-10}$ \cite{Ryan}, 
an observationally 
determined value of the lithium abundance in the atmospheres of hot Population II halo stars. 
It is unclear whether stellar depletion of Li could account for 
such a large deficit, and speculations of non-standard physics being behind the discrepancy 
flourished (see, {\em e.g.} reviews \cite{reviews}). Most recently, 
this lithium problem has been further complicated by the observation of the deterioration of 
the plateau at the lowest metallicities, $Z<1.5\times 10^{-3}$, where the discrepancy with SBBN  value becomes even larger \cite{Sbordone}. 
This might be pointing towards additional "missing" pieces of physics unrelated to the stellar physics, starting from
evolution of primordial gas leading to the formation of PopII stars with lowest metallicities. 

It is important to realize that the observations of lithium abundance along the Spite plateau reflect the ``local'' formation environments of the oldest stars in our Galaxy, while the SBBN predictions are ``global''. 
Non-standard cosmology with an $O(1)$ downward fluctuation of baryon-to-photon ratio in the patch of the 
Universe that includes Milky Way can give $[\lisv/\hydr]_{\rm local} <[\lisv/\hydr]_{\rm SBBN}$ \cite{NollettSA}. 
However, the {\em standard} physics processes  may also lead to the local under- or over-abundance of lithium 
relative to the SBBN prediction. In standard cosmology, the spatial fluctuations of $\lisv/\hydr$ 
are initially small, 
 but consequently amplified by the growth of structure in combination with {\em diffusional 
processes} in the Early and Late Universe. 

The existing astrophysical literature covers the diffusion of elements in stellar atmospheres and in
 the clusters of galaxies \cite{stars,CN,clusters}.  In contrast, the studies of 
primordial element diffusion in the early Universe are very sparse. Ref.  \cite{ML}
addresses the evolution of elemental abundance in the 
linear regime, $\delta \rho/\rho \ll 1$. Although the linear regime by definition 
does not allow for large effects in the abundances, \cite{ML} find that qualitative trend is such that lithium, owing to its larger mass, tends to accumulate more in the minima of gravitational potentials compared to hydrogen. Since the star formation should also occur inside gravitational wells, the qualitative trend inferred from \cite{ML} is $[\lisv/\hydr]_{\rm local} >[\lisv/\hydr]_{\rm SBBN}$, which does not help to solve the lithium problem in any way. 
There is, however,  an important assumption made about the neutrality of 
lithium in \cite{ML}, which does not hold in the 
early Universe. In fact, after the H recombination, lithium exists predominantly in the singly-ionized state,  \lip\ \cite{SH}. There are two reasons for that: firstly, the 5.39 eV ionization potential for lithium means that the recombination temperature is smaller than that of H by the factor of $\sim 2.5$, at which time the density 
of the free electrons is depleted, 
and the recombination rates for Li are less than the Hubble expansion rate. In addition, the non-thermal population of photons from residual $e-p$ recombination causes photo-ionization of neutral Li fraction and keeps its abundance below a few percent level throughout the cosmic history all the way to reionization at $z\sim 10$ \cite{SH}. 

In this paper, we show that the fact that lithium remains in the 
\lip\ state has direct consequences for its diffusion after hydrogen recombination. 
In particular, we show that owing to the large scattering cross section on protons, \lip\ stays spatially 
bound to \hp, and the direction of their diffusion is {\em against} 
the gravitational force in the rest frame where the neutral 
hydrogen. 
From our analysis
it follows that $[\lisv/\hydr]_{\rm grav~min} <[\lisv/\hydr]_{\rm SBBN}$, which can have significant implications for the 
cosmological lithium problem. In the rest of the paper, we expand on this observation in some detail.

{\em Direction of Lithium Diffusion.}
We consider the equations for cosmological fluids of different primordial species with number densities $n_a$, where $a$ spans H, \hef, $e$, $p$, \lisv. Note that after most of the hydrogen becomes neutral, the recombination 
rate for residual $e$ and $p$ is much smaller than the Hubble expansion rate and they can be treated as separate species. 
The presence of small quantities of D and \het\ will not affect the evolution of \lisv.
Thus, we have  the system of equations for the average velocities $V_a$ of individual species,
\begin{eqnarray}
\label{Euler}
\frac{\partial {\bf V}_a}{\partial t} \simeq {\bf g} -\frac{\nabla P_a}{\rho_a} +\frac{q_a}{m_a} {\bf E}
-\sum_b  
\frac{{\bf V}_a-{\bf V}_b}{\tau_{ab}} +\fr{{\bf f}^{\rm ext}_{a}}{m_a}.
\end{eqnarray} 
In these equations, ${\bf g}$ is the gravitational acceleration, ${\bf E}$ is the electric field strength, $P_a$, $\rho_a
= m_a n_a$, and $q_a$ are the partial pressure, mass density, and the electric charge for different species respectively. 
For the electromagnetic effects we assume the tight charge coupling approximation. The ${\bf V}_a-{\bf V}_b$ diffusion term
is governed by the diffusion coefficients $\tau_{ab}^{-1}$, 
\begin{equation}
\label{tau} 
{\tau_{ab}}^{-1} = ({3Tm_b})^{-1}\times {\mu_{ab}^2n_b\langle \sigma_{ab}v^3\rangle},
\end{equation}
that are in turn determined by the transport cross sections $\sigma_{ab}$ 
averaged over the microscopic velocity distribution. We use lower  and upper case 
to distinguish between thermal $v$ and diffusional  ${\bf V}$ velocities.
$\mu_{ab}$ is 
reduced mass and  $T$ is the temperature of the matter species.
Finally, the last term in (\ref{Euler}) accounts for the possibility of additional external forces, such as radiation pressure, Lorentz force, etc., with dependence on species index $a$. 
However, we take ${\bf f}^{\rm ext}_{a}=0$ for the rest of this analysis. 

In the next step, we solve Equations (\ref{Euler}) in the regime of small density perturbations, $\delta \rho_a/\rho_a\la 1$,
and specifically  consider a sub-Jeans regime for baryons, that are forced inside an already formed dark matter halo by its gravitational acceleration ${\bf g}$. For a realistic choice of parameters 
 $1/\tau\to \infty$ is a good zeroth order approximation, 
leading to vanishing ${\bf V}_i$ in hydrostatic equilibrium. 
Thus, assuming that initial distribution of elements is uniform, one gets a relation between gradients of individual pressure contributions and 
${\bf g}$ (see, {\em e.g.} \cite{CN}), 
\begin{equation}
\label{uniform} 
\frac{\nabla P_a}{\rho_a} = \frac{\Sigma n_b m_b}{m_a\Sigma n_b} 
{\bf g} = \frac{\overline{m} }{m_a} {\bf g},
\end{equation}
and the quasi-static version of Eq. (\ref{Euler}) reduces to a set of algebraic equations, 
\be
 {\bf g}\left( 1- \frac{\overline{m} }{m_a}\right)  +\frac{q_a}{m_a} {\bf E} -\sum_b  
\frac{{\bf V}_a-{\bf V}_b}{\tau_{ab}} =0.\label{algebraic}
\ee
We assume 
25\% mass fraction of \hef\, so that $n_{\hef}/n_{\rm H}=1/12$ and 
$
\overline{m} \simeq (4m_p + 12m_p)/(1+12)=\frac{16}{13} m_p.
$
The  reduction of \mav\ due to ionized fraction can be safely neglected. 

For the two dominant neutral components, hydrogen and helium, the solution is readily found:
\be
\label{VHeH}
\frac{ {\bf V}_{\rm He}-{\bf V}_{\rm H}}{\tau_{\rm He H}} ={\bf g}\left(  1-\frac{\overline{m}}{m_{\rm He}} \right)=\frac{9}{13}{\bf g}.
\ee 
One can see that  ${\bf V}_{\rm He}-{\bf V}_{\rm H}$ is parallel to ${\bf g}$, as expected. 

We now turn to the diffusion of charged particles and account for ${\bf E}$ in the equations. 
Solving them for electrons with the use of (\ref{uniform}) and  $m_e\ll m_{\rm atom}$,  one can easily find
$
{\bf E} \simeq -\nabla P_e /(e n_e) =  -(\overline{m}/e){\bf g}_{\rm}
$, where $e$ is the positron charge. 
Carrying this to the equation for protons (or H$^+$), 
we get the solution for the relative diffusion velocity:
\begin{eqnarray}
({\bf V}_{ p} - {\bf V}_{\rm H})
\times\left( \fr{1}{\tau_{p\rm H}}+\fr{1}{\tau_{p\rm He}} \right)
\nonumber
=
-{\bf g}_{\rm } \left[  
\frac{\overline{m}}{m_p/2} -1
\right.
\\
\left.-\left( 1-\frac{\overline{m}}{m_{\rm He}} \right) \frac{\tau_{\rm HeH}}{\tau_{p\rm He}}\right] = -{\bf g}_{\rm } \left[
\frac{19}{13} - \frac{9\tau_{\rm HeH}}{13\tau_{p\rm He}}\right],\label{pH}
\end{eqnarray}
using Equation (\ref{VHeH}).
The appearance of $m_p/2$ in this equation is easy to interpret: the effect of the 
EM force is such that the motion of $e$ and $p$ is tightly 
coupled together, so that their effective mass per particle is $m_p/2$, and indeed lighter than $\mav$. 
This results in the diffusion of both $e$ and H$^+$ { against} the direction of the gravitational 
acceleration, if helium contribution is negligible. 

We are now ready to include the diffusion of Lithium, using already found solutions for 
${\bf V}_p-{\bf V}_{\rm H}$ and ${\bf V}_{\rm He}-{\bf V}_{\rm H}$. The general expression is given 
by 
\begin{eqnarray}
\nonumber
({\bf V}_{\rm Li} - {\bf V}_{\rm H})\times
\left( \frac{1}{\tau_{\rm LiH}}+\fr{1}{\tau_{\rm LiHe}}+\frac{1}{\tau_{{\rm Li}p}} \right) \\
=-{\bf g} \left\{ 
\fr{2\mav}{m_{\rm Li}}-1 -\left(1-\frac{\mav}{m_{\rm He}}   \right) \frac{\tau_{\rm HeH}}{\tau_{\rm LiHe}}
\right.
\nonumber\\
\left.  +\left[ \fr{2\mav}{m_p}-1-\left( 1-\frac{\overline{m}}{m_{\rm He}} \right)\frac{\tau_{\rm HeH}}{\tau_{p\rm He}}\right] 
\frac{\tau^{-1}_{\rm Lip}}{(\tau^{-1}_{p\rm H}+\tau^{-1}_{p\rm He})} \right\}.\label{LiH}
\end{eqnarray}

It turns out that, to a good approximation, we can neglect 
the helium contribution, $n_{\rm He}/n_{\rm H} \to 0$, thus 
$\tau_{p\rm He}^{-1}\to0$, $\overline{m}\to m_p$, and  the cumbersome expressions in Equations (\ref{pH}-\ref{LiH}) simplify to
\begin{eqnarray}
n_{\rm He}/n_{\rm H} \to 0~{\rm limit}:~~~
\frac{ {\bf V}_p-{\bf V}_{\rm H}}{\tau_{p\rm H}} =-{\bf g};
\nonumber
\\
({\bf V}_{\rm Li} - {\bf V}_{\rm H})
\left( \frac{1}{\tau_{\rm LiH}}+\frac{1}{\tau_{{\rm Li}p}} \right)
= {\bf g}_{\rm } \left(  
\frac{5}{7} - \frac{\tau_{p\rm H}}{\tau_{{\rm Li}p}}\right),
\label{Lisimp}
\ea
where in the last formula we approximated $m_{\rm Li} =7 m_p$.
The direction of the lithium diffusional velocity is far from obvious: it depends on the 
competition of the two terms on the r.h.s. of Equation (\ref{Lisimp}), and {\em if} the friction relative to 
$p$ wins ({\em i.e.} small $\tau_{{\rm Li}p}$ limit), the motion of Li$^+$ ions will trace the motion of 
ionized fraction of hydrogen gas. Indeed, Eq. (\ref{Lisimp}) reduces to 
$({\bf V}_{\rm Li} - {\bf V}_{\rm H})/\tau_{p\rm H} = -{\bf g}$, or ${\bf V}_{\rm Li}=
{\bf V}_{ p}$, if $\tau_{{\rm Li}p}^{-1}$ 
is the largest parameter.
We now need additional input with actual size of  $\tau_{ab}^{-1}$.

The scattering of \hef\ on $p$ has been calculated in \cite{Dalgarno}. The value of the 
transport cross section in the range of energies we are interested in, and its weighted average over the Maxwellian velocity distribution is given by 
\be
\sigma_{\rm HeH} \simeq 100 a_B^2;~~\langle \sigma_{\rm He H} v^3 \rangle \simeq (32/\pi)^{1/2}\left( \frac{T}{\mu_{14}} \right)^{3/2} \sigma_{\rm HeH},
\ee
where $T$ is the temperature of the baryonic fluid, $a_B = 1/(\alpha m_e)$, is the Bohr radius, and $\mu_{14} = 4m_p/5$. 
The cross sections of a singly-charged ion on a neutral atom can be approximated  as $\sigma_{ab} \simeq 2.2 \pi \left( \frac{\alpha_{\rm pol}(b) \alpha}{2E}\right)^{1/2}$ \cite{MM}, leading to 
\begin{eqnarray}
\label{MMsc}
\langle \sigma_{ab}v^3  \rangle \simeq 20\pi a_B^2 \frac{{\rm Ry}^{1/2}T}{\mu_{ab}^{3/2}}~
\left(  \fr{\alpha_{\rm pol}(b)}{\alpha_{\rm pol} (\rm H)} \right)^{1/2},
\end{eqnarray}
where $\alpha_{\rm pol}(b)$ is the atomic polarizability of the neutral 
species $b$: $\alpha_{\rm pol}(\hydr) =\fr92 a_B^3$ and
$\alpha_{\rm pol}({\rm He}) = 1.38 a_B^3$. Ry stands for the hydrogen binding energy, ${\rm Ry}\equiv \alpha^2m_e/2\simeq 13.6$ eV. We should note that $p$-H scattering is in practice a more complicated process due to the identical nature of the nuclei involved, and a far more elaborate treatment of the $p$-H cross section can be found in  \cite{pH}. However, for the accuracy of our discussion, we shall still approximate it with Eq. (\ref{MMsc}).
Finally, and most importantly, the $p$-\lip\ scattering is given by the Rutherford formula, 
\be
\label{Ruth}
\langle \sigma_{{\rm Li} p} v^3\rangle = \frac{8\pi^{1/2}\alpha^2}{(2T)^{1/2}\mu_{17}^{3/2}}
\times \ln \Lambda
=16\sqrt{2\pi}a_B^2\frac{\rm Ry^2\ln\Lambda}{T^{1/2}\mu_{17}^{3/2}},
\ee
where $\mu_{17} = \fr78 m_p$, and $\ln \Lambda$ is the  Coulomb logarithm. For 
the conditions of primoridal 
plasma after the recombination, its value is large, $\ln \Lambda \sim 40$, and weakly dependent on temperature. Because of the long-range nature of the EM force, Eq. (\ref{Ruth}) exhibits strong 
enhancement by $({\rm Ry}/T)^{3/2}\ln \Lambda$ at small velocities/low temperatures.

We are now ready to determine the sign 
of the r.h.s. bracket in the simplified formula (\ref{Lisimp}):
\begin{eqnarray}
\frac57 - \frac{\tau_{p\rm H}}{\tau_{{\rm Li}p}}
\nonumber
\sim \frac57 - 300 \times \frac{X_e}{10^{-3}} \times \frac{\ln \Lambda}{40}
\times \left( \frac{0.01\rm eV}{T_{\rm baryon}}\right)^{3/2}<0,
\\
\label{bracket}
{\bf V}_{\rm Li} - {\bf V}_{\rm H} \propto - {\bf g}_{\rm }, ~~~~~~~~~~~
\end{eqnarray}
where the abundance of free protons in the primordial plasma is 
the same as the electron ionization fraction $X_e$.
It is easy to see that for the cosmological parameters between recombination and re-ionization, 
the expression (\ref{bracket}) is negative. In Figure (\ref{fig:levels}), we plot the the positivity condition on $X_e$-redshift plane, assuming standard relations between $T$, photon temperature $T_\gamma$ and redshift $z$. 
The separatrix stays firmly below cosmological $X_e(T)$ at all redshifts.
One can see that even tiny values of 
$X_e$ would lead to a tight coupling of lithium to H$^+$, resulting in outward diffusion of lithium.
We can also quantify the ratio of relative velocities,
\be
\label{comparison}
\frac{|{\bf V}_{\rm Li} - {\bf V}_{\rm H}|}{|{\bf V}_{ p} - {\bf V}_{\rm H} |}\simeq 
 1 - \fr{12}{7}\frac{\tau_{{\rm Li}p}}{\tau_{p\rm H}}  = 1 - {\cal O}(10^{-2}),
\ee
which are different only at ${\cal O}(\%)$ level. Conclusions of Eqs. (\ref{bracket}) and (\ref{comparison}) are due to 
large size of the Rutherford cross section that overcomes the rarity of H$^+$. Inclusion of He into this analysis does 
change the conclusions somewhat: 
while \lip\ remains tightly bound to H$^+$, the outward diffusion 
of ionized H$^+$ is no longer guaranteed. We find that for most of the redshift  of interest, $z>30$, both 
$p$ and Li diffuse out of overdensities not changing the qualitative details of the simplified analysis.

\begin{figure}
\includegraphics[width=\linewidth]{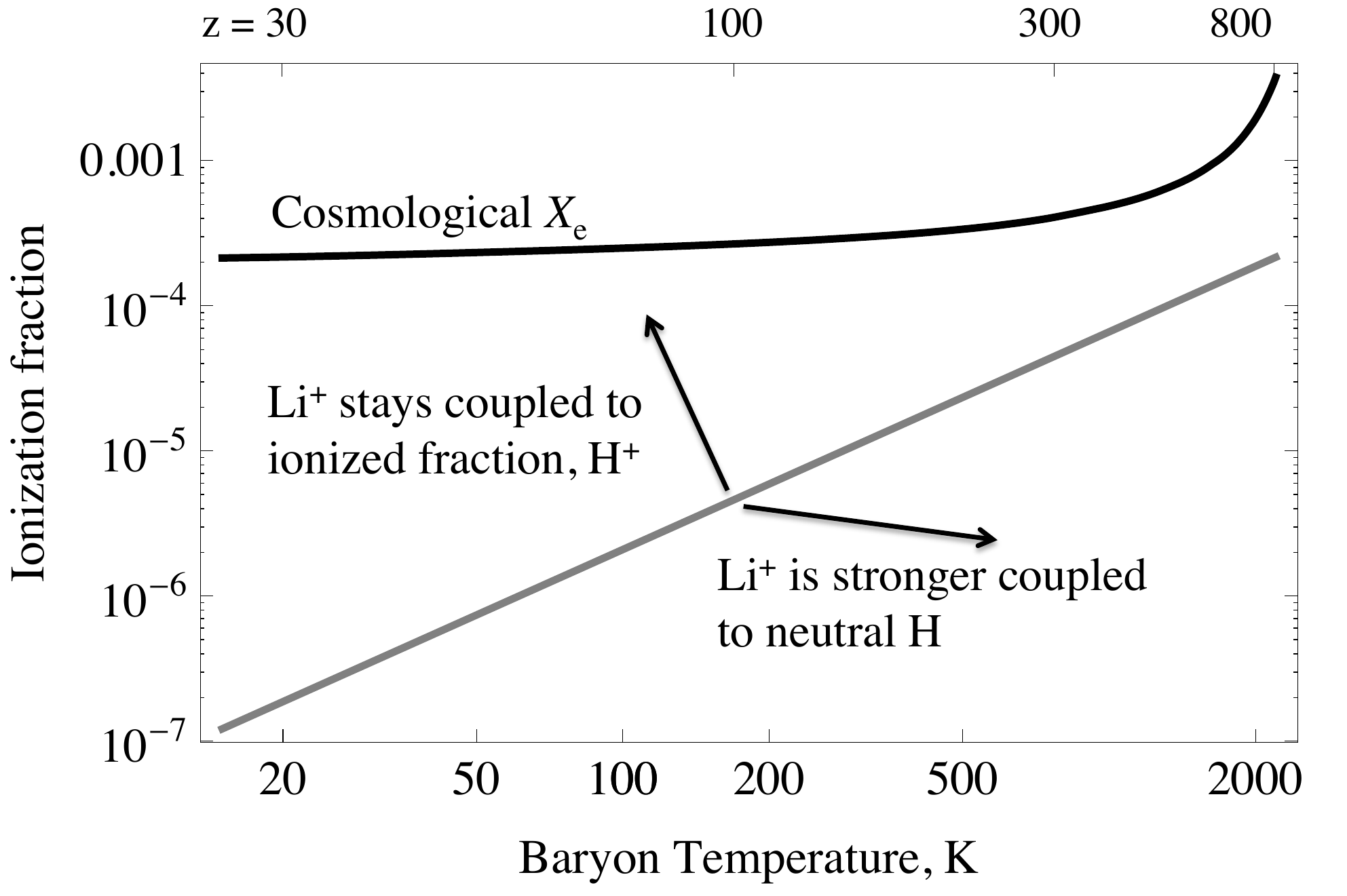}
\caption{Post-recombination ionization fraction $X_e(T)$ (black line) and the separatrix for the direction of 
${\bf V}_{\rm Li} - {\bf V}_{\rm H} $ relative to ${\bf g}$. 
Above the gray line Eq. (\ref{bracket}) holds, and since it stays always below the black curve, lithium diffuses ``out'', leading to $[\lisv/\hydr]_{\rm grav~min} <[\lisv/\hydr]_{\rm SBBN}$.}
\label{fig:levels}
\end{figure}

{\em  Do we understand the magnitude and sign of possible 
$[\lisv\rm /H]$ variations? }
While we have shown that, quite unexpectedly, the direction of lithium diffusion in the early Universe 
after the recombination is against local gravitational force, it is 
clear that it would be difficult to create variations in 
 lithium abundance at $O(1)$ level. Let us assume to good accuracy that $1/\tau_{{\rm Li}p}$ is the largest coefficient, so that  motion
of \lip\ and H$^+$ are spatially linked. Then, using the continuity equations we can tie the variation in lithium 
abundance that develops at redshift $z_f$ to the local halo density,

\begin{eqnarray}
\frac{[{\rm Li/H}]_{\rm SBBN}-[{\rm Li/H}]_{\rm halo} }{[{\rm Li/H}]_{\rm SBBN}} \simeq  -\int_{t_i}^{t_f} dt \frac{\nabla \cdot {\bf g} }{1/\tau_{p\rm H}}~~~~~~~~~~
\\
\label{final}
=\int_{t_i}^{t_f} dt~~ 4\pi G_N\rho_{\rm halo} \tau_{p\rm H} 
=\fr32\int^{z_i}_{z_f} \fr{dz}{z+1}~\frac{\rho_{\rm halo}}{\overline{\rho}(z)} H(z)\tau_{p\rm H}
\nonumber\\\nonumber
\simeq 1 \times 10^{-2} \left( \Delta \over 200\right) \left( 1+z_f \over 20 \right)^{-3/2}.~~~~~~~~~~~~
\ea
Here  $\overline\rho$ is the average matter density, while
$\rho_{\rm halo}$ is the mass density of the halo, which is 
presumably contributed to mostly by the cold dark matter in the sub-Jeans regime. 
The Hubble expansion rate, $H(z)$, makes appearance in this formula. 
Notice that while $\overline{\rho}(z)$, $H(z)$, and $\tau_{p\rm H}$ depend on time/redshift in a simple calculable way, 
$\rho_{\rm halo}$ can vary by orders of magnitude, depending on the position with respect to dark matter haloes.
The last line of Eq. (\ref{final}) assumes typical overdensity of $\Delta \equiv \rho_{\rm halo}/\overline{\rho}(z) \sim 200$ for collapsed haloes, and ignores overdensity of neutral hydrogen, 
as expected for idealized sub-Jeans mass haloes. 
We have confirmed this estimate by direct numerical integration of system (\ref{algebraic}).

Judging by these results, one would conclude that the post-recombination diffusion-induced variation 
of lithium abundance is going to be small {\em on average}. That does not 
mean, however, that such variations will be small  everywhere  in the Universe (see  Fig. \ref{fig:profile}). The central 
regions of the halos, which have the shortest cooling time and thus most likely to form the first stars, can have significantly higher densities, and thus 
create ${\cal O}(1)$ depletion of lithium abundance in a spatially small patch. Fig. (\ref{fig:profile}) may also help explain why the Lithium depletion is larger in lowest metallicity stars \cite{Sbordone}, as they must have formed earlier at the centres of more massive haloes (with denser cores and thus shorter cooling times), which had experienced a higher level of depletion. 
Therefore, in light of finding a physical mechanism that can 
lead to the depletion of Li in overdensities, we may 
re-state the lithium problem in the following way: 
in addition to asking ``how Li got destroyed?'', one could also question ``how likely is that our observations sample a special 
part of the Universe, where Li was depleted before stars formed?''. 

\begin{figure}
\includegraphics[width=\linewidth]{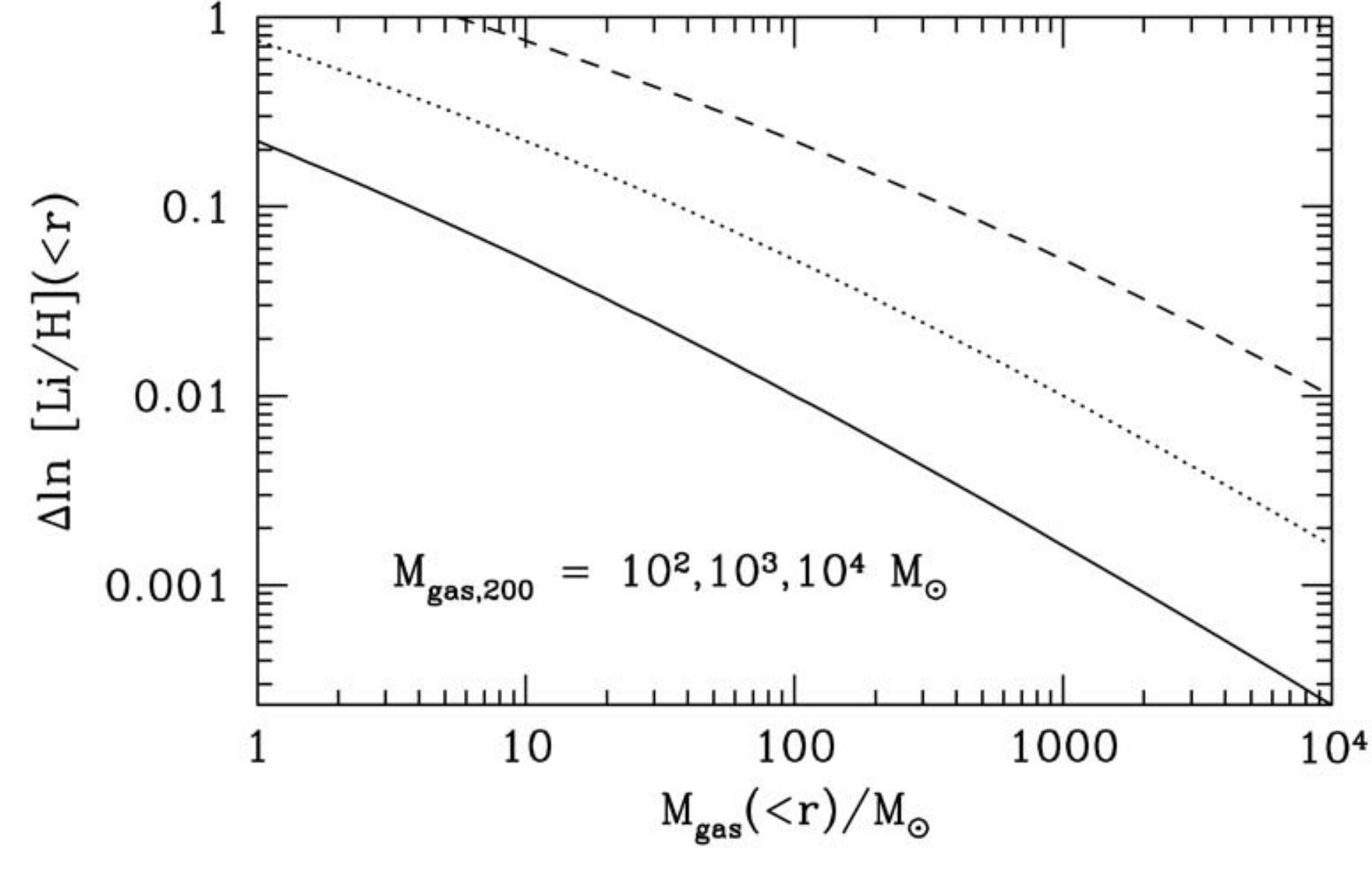}
\caption{Level of lithium depletion within primordial dark matter haloes at $z= 20$, with virial gas masses of $10^2$ (solid), $10^3$ (dotted) and $10^4$ M$_{\odot}$ (dashed). We assume an NFW halo \cite{Navarro:1996gj} with concentration of $c=5$ for dark matter, and ignore gas overdensity, as expected in the sub-Jeans mass regime.  }
\label{fig:profile}
\end{figure}

We shall briefly emphasize the limitations of our approach: As the halo mass approaches the Jeans mass, the gas overdensity will become significant, which in turn suppresses $\tau_{p\rm H}$ in Eq. (\ref{final}), and thus the depletion rate. Supersonic halo mergers lead to turbulence which further suppresses the diffusion. Furthermore,  the ionization fractions may differ from the cosmological values at significant gas overdensities. On the other end, for smaller haloes the supersonic relative velocities of dark matter and baryons suppresses diffusion, except within a small fraction of cosmic volume \cite{Tseliakhovich:2010bj}. Thus, only careful hydrodynamic simulations of early star formation, which include diffusion effects, will be able to confirm the real(istic) magnitude of lithium depletion in population II and III stars.  

Finally, we would like to stress that perhaps our most interesting and novel  result is that lithium 
remains closely tied to ionized fraction of the gas. Therefore, any {\em additional} forces ${\bf f }_a$ 
that act on neutral and charged components differentially, could play a role in creating variations in lithium abundance.
Interesting candidates for creating such forces are radiation pressure/stellar winds from first stars, and
possibly primordial magnetic fields. This whole scope of issues 
deserves close attention due to the continuing interest in the lithium problem.


We would like to thank Dr. J. Pradler for helpful discussions. 

\end{document}